\documentclass{article}

\usepackage[dvips]{graphicx}

\renewcommand{\abstractname}{}
\renewcommand{\refname}{Reference:}

\title{To the modification of methods of nuclear chronometry in astrophysics and geophysics}

\author{V.~S.~Olkhovsky\thanks{E-mail: olkhovsk@kinr.kiev.ua}, M.~E.~Dolinska \\
\small\emph{Institute for Nuclear Research, National Academy of Sciences of Ukraine} \\
\small\emph{prosp. Nauki, 47, Kiev-28, 03680, Ukraine}}

\date{\small\today}

\begin{document}

\maketitle

\renewcommand{\abstractname}{}
\begin{abstract}
In practically all known till now methods of nuclear chronometry there were usually taken into account the life-times of only fundamental states of $\alpha$-radioactive nuclei. But in the processes of nuclear synthesis in stars and under the influence of the constant cosmic radiation on surfaces of planets the excitations of the $\alpha$-radioactive nuclei are going on. Between them there are the states with the excited $\alpha$-particles inside the parent nuclei and so with much smaller life-times. And inside the large masses of stellar, terrestrial and meteoric substances the transitions between different internal conditions of radioactive nuclei are accompanied by infinite chains of the $\gamma$-radiations with the subsequent $\gamma$-absorptions, the further $\gamma$-radiations etc. For the description of the $\alpha$-decay evolution with considering of such excited states and multiple $\gamma$-radiations and $\gamma$-absorptions inside stars and under the influence of the cosmic radiation on the earth surface we present the quantum-mechanical approach, which is based on the generalized Krylov-Fock theorem.

Some simple estimations are also presented. They bring to the conclusion that the usual (non-corrected) ``nuclear  clocks'' do really indicate not to realistic values but to the \emph{upper limits} of the durations of  the $\alpha$-decay stellar and planet  processes.
\end{abstract}



\section{Introduction
\label{sec.introduction}}

Usually in the practical applications of the standard methods of nuclear chronometry for the stellar and terrestrial processes the decays of nuclei-chronometers from only ground states are taken into account. However, during $s-$ and $r-$ processes of nuclear synthesis in stars and supernova not only ground states but also the excited states of synthesized nuclei are being formed [1]. It follows directly from the Geiger and Nutall law for the $\alpha$-decay that the lifetime $\tau_{exc}$ of such excited $\alpha$-decaying nucleus very strongly depends on the $\alpha$-particle kinetic energy. In many cases the lifetime $\tau_{exc}$ is diminished by several orders with the increasing of the $\alpha$-particle energy by 1-2 MeV! But up to now no systematic experimental study of the dependence of lifetimes of the excited radioactive nuclei relative to the $\alpha$-decays on their excitation energy had been undertaken because of the difficulties connected with much more rapid (within $10^{-13}$ -- $10^{-9}$ sec) and strong $\gamma$-decays of excited nuclei and so previously it had assumed that there is no practical reason to take into account the much more slow processes of the $\alpha$-decays from these states. Namely in [2,3] it had been firstly supposed that inside large masses of stellar substance a part of radioactive nuclei could be supported in the excited states during a long time for rather wide temperature interval due to chains of subsequent emissions and quasi-resonance absorptions of $\gamma$-quanta by nuclei-chronometers when the energy losses, caused by recoils during emitting and absorbing, are compensated by the nucleus thermal kinetic energy.

And also the formation of the excited states of the $\alpha$-radioactive nuclei in the terrestrial surface layers and in the meteorites is present under the influence of the weak but constant cosmic radiation.

Further, it is experimentally shown in [4] that the lifetime of the $\beta$ - radioactive isotope $^{187}{\rm Re}$ in bare (without electrons) atom is less $10^{9}$ times than the lifetime of the usual atom with totally filled electronic shell (it becomes $\sim$ 30 years instead of $\sim 3 \cdot 10^{10}$ years!). And it is known that inside stars the nuclei partially or totally are being deprived their electronic shells and the stellar matter is in the plasma state with the free nuclei and electrons.

\section{
\label{sec.2}}

Here we shall deal with the $\alpha$-radioactive nuclei-chronometers. In standard nucleo-chronometry techniques, together with the decay function $L(t-t_{0}) = \exp(-\Gamma_{0}(t-t_{0})/ \eta)$ ($\Gamma_{0}$ being the $\alpha$-decay width of the ground state, $t_{0}$ being an initial time) and the surviving function $W(t) = 1 - L(t)$ one uses the abundances $P$ and $D$ of parent and daughter nuclei connected with  $L$  and  $W$  by the following evident relations:
\begin{equation}
\begin{array}{cc}
  L(t-t_{0})=P(t) / P(t_{0}), &
  W(t-t_{0})=[P(t) - P(t_{0})] / P(t_{0})
\end{array}
\label{eq.1}
\end{equation}
and
\begin{equation}
  P(t) + D(t) = P(t_{0}) + D(t_{0}),
\label{eq.2}
\end{equation}
or
\begin{equation}
  D(t)-D(t_{0})=P(t_{0})W(t-t_{0}) = P(t)[ \exp(\Gamma_{0}(t-t_{0})/\eta)-1].
\label{eq.2a}
\end{equation}

Usually equality (\ref{eq.2a}) is divided by the abundance of \emph{another} stable isotope $D_{x}$ which does not obtain any contribution from the decay of the parent nuclei (i.~e. $D_{x}$  does not depend on time). And as a result, (\ref{eq.2a}) acquires the following form
\begin{equation}
  p(t) [\exp(\Gamma_{0}(t-t_{0})/)-1] - d(t) + d(t_{0}) = 0
\label{eq.2b}
\end{equation}
where $p=P/D_{x}$  and  $d=D/D_{x}$ . Measuring  $p=P/D_{x}$  and  $d=D/D_{x}$ in different samples (or in different separate parts of the same sample), we obtain the plot (\ref{eq.2b}) at the plane  $p=P/D_{x}$ , $d=D/D_{x}$  which has the form of a straight line, the slope of which with respect to axis p  permits simply to define age  $t-t_{0}$ (see, for instance, [1]).

For further describing the $\alpha$-decay evolution inside stars and under the cosmic radiation on the earth surface we shall use the Krylov-Fock theorem [5] (see also its generalization in [6-9]):
\begin{equation}
L(t-t_{0}) =  |f(t-t_{0})|^{2} / |f(t_{0})|^{2}
\label{eq.3}
\end{equation}
where
\begin{equation}
f(t-t_{0}) = \int\limits_{0}^{+\infty} |G(E)|^2 \exp [-iE( t-t_{0})/\eta] \; dE
\label{eq.4}
\end{equation}
is the characteristic function for the energy $(E)$ distribution in a decaying state with the weight amplitude $G(E)$. The decay rate per a time unit is [10]
\begin{equation}
\rho(t-t_{0}) = d [1 - L(t-t_{0})]/ d(t-t_{0}) .
\label{eq.5}
\end{equation}

When the decay is going on by several channels (for example, by $\alpha$ - and $\gamma$ -decays from the first excited state of the parent nucleus) instead of (\ref{eq.4}) we have [10]
\begin{equation}
\rho_{1i}(t-t_{0}) = (\Gamma_{i}^{1}/\Gamma_{1})d[1 - L_{1}(t-t_{0})]/ d(t-t_{0})=(\Gamma_{i}^{1}/\eta)\exp (-
\Gamma_{1}(t - t_{0} )/\eta )
\label{eq.5a}
\end{equation}
and  the probability to be decayed by the $i$-th channel is described by
\[
  W_{1i}(t-t_{0}) =\int\limits_{t_{0}}^{t}( \rho_{1i} (t') dt' =  (\Gamma_{i}^{1}/\Gamma_{1} [1 - L_{1}(t-t_{0})].
\]

The decay of an ensemble of radioactive nuclei can go on simultaneously with its preparation (in particularly, with the nucleosynthesis or with decays from the previous state). For the last cases one can use the following almost evidently generalized (see, for instance, [7-9]) expression for the decay rate per a time unit:
\begin{equation}
  I(t-t_{0}) = \int\limits_{t_{0}}^{t} \rho_{1}^{0}\: (t')\: \rho_{0}^{1}\: (t- t')\; dt'
\label{eq.5b}
\end{equation}
with  $\rho_{m}^{(n)}\:(t) = [\delta_{m1}\:\Gamma_{\gamma}^{1}/\Gamma_{1} + \delta_{m0}]\; d[1 - L_{m}^{(n)}\:(t)]/\; dt$ $(m \neq n = 0, 1)$, where decay functions $L_{1}^{(0)} (t)$ and $L_{0}^{(1)} (t)$ characterize the decay of an \emph{initial} (the first excited) state and the \emph{subsequent} (ground) state, formed after the $\gamma$ -decay of that initial one, respectively; $\Gamma_{1}  = \Gamma_{\alpha}^{1} + \Gamma_{\gamma}^{1}$ , $\Gamma_{\alpha}^{1}$ and $\Gamma_{\gamma}^{1}$ being the $\alpha$-decay and $\gamma$ -decay widths of the excited parent $\alpha$-radioactive nucleus (with the first excited state of the internal $\alpha$-particle).  Functions $L_{1}^{0} (t)$ and $L_{0}^{1} (t)$ are defined by the appropriate spectra of energy distributions $|G_{1}^{(0)}(E)|^{2}  =  const / [(E_{1} - E)^{2}  + \Gamma_{1}\:^{2} / 4 ]^{-1}$  and  $|G_{0}^{(1)}(E)|^{2} =  const / [(E_{1} - E)^{2}  + \Gamma_{1}\:^{2} / 4]^{-1}\cdot [(E_{0} - E)^{2}  + \Gamma_{0}\:^{2}/ 4]^{-1}$ , respectively. Here $E_{0}$ , $E_{1}$  and E denote the mean (resonance or level) energy of the ground and the first excited state and the factual energy in the system of the internal motion of a parent nucleus, respectively. At the usual approximation of very small widths the lower limit 0 of the integral in (\ref{eq.4}) can be taken as $-\propto$  (see, for instance, [6-10] and references therein) and the calculation of decay functions $L_{1}^{(0)}(t)$ and $L_{0}^{(1)}(t)$ gives an exponential form for $L_{1}^{(0)}(t)$, i.~e. $L_{1}^{(0)}(t-t_{0}) = L_{1}(t-t_{0}) = \exp[-\Gamma_{1}(t-t_{0})/\eta]$, and at the additional approximation of very small quantities $\Gamma_{1}/(E_{1} -E_{0})$ and $\Gamma_{1} /\Gamma_{0}$   also an exponential form for $L_{0}^{(1)} (t-t_{0})$, i.e. $L_{0}^{(1)}(t-t_{0}) = L_{0}^{(0)}(t-t_{0}) = \exp[-\Gamma_{0}(t-t_{0})/\eta]$.

As it was shown in [2,3], the evaluation of the probability of the parent nuclei to be decayed (and consequently the daughter nuclei to be formed) even with the single absorptions of emitted ( $\gamma$-quanta with the subsequent $\gamma$ - and $\alpha$ -decays  is being much more bulky. In this evaluation we have to use the expression  $M(t)= q M_{0} [1 - \exp(- ct/\mu]$ (with c,$\mu$ and $q$ being the light velocity, the inversed $\gamma$ -quanta absorption coefficient inside the matter and the dimensionless quantity, which depends on small loss of the $\gamma$ -quantum energy due to the $\gamma$ -scattering by nuclei and by electrons, respectively) for the $\gamma$ -quanta propagation inside the matter sample and the relation $\varepsilon_{\gamma} = E_{1} -E_{0}  - \varepsilon_{recoil}  + D$  (with $\varepsilon_{recoil}$  and {D} being the recoil kinetic energy of the nucleus after the $\gamma$ -quantum emission or absorption and the Doppler width for the resonant $\gamma$-emission and absorption, respectively). After a series of simplifications at the additional approximations  $\varepsilon_{recoil}\sim(E_{1} -E_{0})^{2} / 2Am_{n}c^{2}$  ($A$ and $m_{n}$ being the atomic number of parent nucleus and the mean nucleon mass, respectively),  $D\sim2[\varepsilon_{recoil} kT]^{1/ 2} $  ({k} and {T} being the Boltzman constant and the sample temperature, respectively),  $\Gamma_{0} <<\Gamma_{1} , c\eta/\mu$  and  $t >> \eta/\Gamma_{1}, \mu/c$  and also considering the diminution of \emph{directly} decaying parent nuclei due to the $\gamma$ -absorption,  it was obtained the following expression:
\begin{equation}
 D(t) = D(t_{0}) + [P_{0}(t_{0}) + P_{1}(t_{0}) [1 - \exp(-\Gamma_{0}(t - t_{0} )/\eta)] + P_{1}(t_{0})Q_\lambda[1 - \exp(-\Gamma_{0}(t - t_{0})/\eta )
\label{eq.6a}
\end{equation}
with
\[
 Q_{\lambda} = (\Gamma_{\alpha}^{1}/\Gamma_{1})qM_{0}[P_{0}(t_{0})+ P_{1}(t_{0})(2\Gamma_{1} + c\eta/\lambda) / 2(\Gamma_{1} + c\eta/\lambda)],
\]
or
\begin{equation}
D(t) = D(t_{0}) + [P_{0}(t_{0}) + (1- Q_{\lambda})P_{1}(t_{0})] [1 - \exp(-\Gamma_{0}(t - t_{0})/\eta)] + Q_{\lambda} P_{1}(t_{0}),
\label{eq.6b}
\end{equation}
$P_{0}$ and $P_{1}$ being the abundances of the ground and the so excited states of the parent $\alpha$-radioactive nuclei, respectively. The tentative estimations of $q$ gave values which are approximately within ($\frac{1}{2}$, 1).

Taking into account not only single (one-step) $\gamma$-absorptions but all possible \emph{multiple} $\gamma$-absorptions, one can evaluate an every step of $\gamma$-absorptions by an equal contribution with $M\cong qM_{0} (2\Gamma_{1} + c\eta/\lambda)/ 2(\Gamma_{1} + c\eta/\lambda)$ and easily sum up them in (\ref{eq.6a}) or (\ref{eq.6b}). And then, if  $\Gamma_{\alpha}^{1}/\Gamma_{1}$ is not very small (let us say, $\geq 0.1$), one can also roughly take $N\Gamma_{\alpha}^{1}/\Gamma_{1}\cong1$ in (\ref{eq.6b}) ($N$ being an effective number of the considered $\gamma$-absorptions steps).  We can confirm such valuations with the help of the simple evident reasoning. For the lifetimes $\tau_{\gamma}=\eta/\Gamma_{\gamma}^{1}$  larger than $10^{-13}$ sec and  $\Gamma_{\gamma}^{1}\ll D$  and, moreover, for $A\cong 250,\;\varepsilon_{\gamma}\cong 50keV$ and $N\cong10$ the quantity $N\varepsilon_{\rm recoil}$ satisfies the following expressions $N\varepsilon_{\rm recoil}< D$ when $T > 300^{\circ}K$ (for terrestrial lumps) and $N \varepsilon_{recoil} \ll D$ when $T \cong 10^{9} \: {}^{\circ}{\rm K}$ (inside stars). Although up to now the partial $\alpha$-decay widths for excited states experimentally are not studied we can expect that the condition $N<10$ is rather realistic on account of the Geiger and Nutall law, at least for high-energy excited states. If the values of $\Gamma_{\gamma}$ and $N\varepsilon_{\rm recoil}$ get into such spreads D we can generalize (\ref{eq.6b}) for multiple $\gamma$-absorptions and write (with $Q_{\lambda}\rightarrow q$)
\begin{equation}
D(t) = D(t_{0}) + [P_{0}(t_{0}) + P_{1}(t_{0})(1- q)] [1 - \exp(-\Gamma_{0}(t - t_{0})/\eta)] + qP_{1}(t_{0})
\label{eq.6c}
\end{equation}
and then at the approximation of $q \rightarrow 1$
\begin{equation}
  D(t) = D(t_{0}) + P_{0}(t_{0})[1 - \exp(-\Gamma_{0}(t - t_{0})/\eta )] + P_{1}(t_{0}).
\label{eq.6d}
\end{equation}

The results (\ref{eq.6c}) and (\ref{eq.6d}) are valid at the approximations of infinitely large medium volumes, and sufficiently large times $t-t_{0}$ (which are much larger than mean lifetimes of excited states and mean times of the free flight of $\gamma$-quanta inside the medium and also than times of quantum oscillations caused by different interference processes described by applying the Krylov-Fock theorem, generalized in [6-9]) and at the condition $(\Gamma_{0} /\Gamma_{\alpha}^{1})N\geq 1$.

We shall analysis of the new relation (\ref{eq.6a}) or (\ref{eq.6c}) in comparison with the previously known equation (\ref{eq.2b}) both of which are basic for the determination of age of a pattern, with and without taking into account the intermediate $\gamma$-absorptions inside the sample matter. For the convenience we rewrite (\ref{eq.2b}), taken for the same conditions as (\ref{eq.6d}), using the same terms as in (\ref{eq.6d}), in the following equivalent form:
\begin{equation}
 D(t) = D(t_{0}) + [P_{0}(t_{0}) + P_{1}(t_{0})] [1 - \exp(-\Gamma_{0}(t - t_{0})/\eta)].
\label{eq.2c}
\end{equation}

From the simple comparison of (\ref{eq.2c}) and (\ref{eq.6d}) one can see that
\begin{itemize}

\item
(i) for the same $D(t_{0})$, $P_{0}(t_{0})$, $P_{1}(t_{0})$, $t_{0}$ and $\Gamma_{0}$ at any moment t the value of \emph{D(t)} in (\ref{eq.6d}) is larger than in (\ref{eq.2c}) by the quantity $P_{1}(t_{0}) \exp(-\Gamma_{0}(t - t_{0})/\eta)$;

\item
(ii) the same value \emph{D(t)} is obtained in (\ref{eq.6d}) at an earlier moment \emph{t} than in (\ref{eq.2c});

\item
(iii) the larger is the contribution of $P_{1}$ into the sum $P_{0} + P_{1}$, the earlier is the moment \emph{t} in (\ref{eq.6d}) in comparison with (\ref{eq.2c}) at which the same value of \emph{D(t)} is obtained.
\end{itemize}

Now we illustrate the inference (ii) by the following instance:
if $P_{0}(t_{0}) = P_{1}(t_{0}) =(1/2)\, P(t_{0})$, then the same value of \emph{D(t)} is obtained for the values of $t=t_{\rm usual}$ in (\ref{eq.2c}) and $t=t_{\rm real}$ in (\ref{eq.6c}) which are connected by the following striking relation
\begin{equation}
t_{real} = t_{usual}  - (\Gamma_{0} /\eta)\:\ln{2}
\label{eq.7}
\end{equation}
(of course, we imply that $t_{\rm usual} - t_{0} >(\Gamma_{0} /\eta)ln2 )$.



\renewcommand{\tablename}{Table}
\begin{table}
\begin{center}
\begin{tabular}{|c|c|c|c|c|c|c|c|c|} \hline
  $t_{real}, years$ &
  $2\cdot 10^{3}$&
  $4\cdot 10^{3}$ &
$6\cdot 10^{3}$ &
$8\cdot 10^{3}$ &
$1\cdot 10^{4}$ &
$1\cdot 10^{6}$ &
$0.8\cdot 10^{8}$ &
$1.8\cdot 10^{9}$ \\ \hline
  $t_{usual}, years \cdot 10^{9}$&
$3.32002 $ &
$3.32004 $ &
$3.32006 $ &
$3.32008$ &
$3.3201$ &
$3.321$ &
$3.4$ &
$3.5$ \\ \hline
\end{tabular}
\end{center}
\caption{The impressive calculation results for the $\alpha$-decay of the nucleus-chronometer $^{238} U$  with the lifetime $4.5 \cdot 10^{9}$ years on the base of (\ref{eq.7}).
\label{table.1}}
\end{table}

So, sometimes billions years obtained by the usual nuclear-chronometry method can correspond to several thousands years. Of course, the results (i) and (ii) will be even stronger if one will consider more than one excited state of the parent nuclei with the excited internal $\alpha$-particles. It is possible to state that the usual (non-corrected) "nuclear clocks" do really indicate to the \emph{upper limits} of the durations of real decay processes.


\section{
\label{sec.3}}

 Now we consider the necessity of revising the methods of the terrestrial nuclear chronometry taking into account the relatively weak but constant cosmic radiation on the upper earth layers up to the depth of  $\sim 5$ meters. Under the influence of the cosmic radiation, three different processes are going on: (1) the constant formation of the excited nuclei-chronometers with much smaller life-times than for the nuclei-chronometers in the ground state, (2) the constant acceleration of the $\alpha$-decay through knocking-out the $\alpha$-particles by cosmic protons and (3) the constant nonzero removal of nuclei-chronometers through the channels of inevitable rearrangement nuclear reactions. \emph{Both processes lead to the real diminishment of the results of measurements of the decay times.}

 Let us analyze the diminishment caused by the first and the second kinds of processes.

 In our case, taking into account the cosmic radiation (supposing its flux to be constant in time) and using the same method of the generalization of the Krylov-Fock theorem as in [6-9], we, instead of $L(t-t_{0})$ (with the consequences (1)-(2) ), obtain:
\begin{equation}
L_{P} (t- t_{0}) = [1 - a \ast (t - t_{0}) ]L(t-t_{0})
\label{eq.8}
\end{equation}
for \emph{a} unit ($1 \: {\rm cm}^{3}$) chronometer volume, where  $a = j_{cosm}\,\sigma\nu n$, $j_{cosm}$ is the cosmic (mainly proton) radiation flux (in ${\rm cm}^{-2}\; {\rm sec}^{-1}$), (is the total cross section of all reaction proton-chronometer processes with the removal of the nuclei-chronometers, $\nu$ is \emph{the number of multiple collisions in the medium} after the first proton-chronometer collision and \emph{n} is the mean nucleus-chronometer number along the 1 cm - depth and $L_{P} (t - t_{0})$ includes \emph{all parent-nucleus diminutions through collisions of chronometers with cosmic protons} . Here for the simplicity we neglect the elastic and inelastic scattering. We see that for $t - t_{0} \geq 1/a$,  $ L_{P \rightarrow D} (t- t_{0}) = 0$  and  $P(t -t_{0}) = 0$ .

If we select only the reactions $(p,p' \alpha)$ with the general cross section $\sigma_{D}$ which strongly accelerate the emission of the $\alpha$-particles and the formation of daughter nuclei, neglecting all other processes, then we obtain the particular relation
\begin{equation}
L_{P \rightarrow D} (t- t_{0}) = [1 - a_{D} \ast (t - t_{0}) ]L(t-t_{0})
\label{eq.8a}
\end{equation}
where $a_{D} = j_{cosm}\sigma_{D}\nu$,  and $L_{P \rightarrow D} (t- t_{0})$ denotes the decay function in this case. And in this case relation (\ref{eq.2b}) passes into
\begin{equation}
p(t) [ [1 - a_{D}\ast(t - t_{0})] \exp (\Gamma_{0}(t-t_{0})/\eta)-1] - d(t) + d(t_{0}) = 0 .
\label{eq.2c}
\end{equation}
From the comparison of (\ref{eq.2c}) with (\ref{eq.2b}) one can see that when $t - t_{0} \rightarrow 1/a$  the time duration defined by the old method (without taking the cosmic radiation into account) becomes very large - much more than $\Gamma_{0}/\eta$.

For qualitative evaluations we have taken $\sigma = 3\star 10^{-25} cm^{2}$  and $\eta$ between $10^{2} \div 10^{3}$ for the mean proton energy $\sim 10^{9}$ eV,  the flux $j_{cosm} = 0.85 (cm^{2} sec)^{-1}$ in the top atmosphere layer  or  $1.75 \ast 10^{-2} (cm^{2} sec)^{-1}$ on the sea level [10] and  $n= 1cm/3 \ast 10^{-8} cm =0.33 \ast 10^{8}$.   Of course, practically it impossible now to calculate ( because  the effective value of ( is defined not only by mean proton energy and by nucleon, cluster and fragment binding energies but also by usually unknown cross sections of all possible reactions for wide energy region - and hence we have used very simplified evaluations. Then we have obtained values of \emph{a} between $1/ 1.5 \ast 10^{8}$ years and $1/ 2.7 \ast 10^{5}$ years.  From the result we can see that when  $P(t) \rightarrow 0$,  for both values of \emph{a} and also in both cases  (with valid and invalid relation (\ref{eq.2c}) ) the real time duration is essentially less than without taking the cosmic radiation into account.

As to the more deep terrestrial layers and the core of the earth, we have to take into account the history of the earth formation. Now there is no unique generally accepted theory or even model of the planet origin. There are two known groups of such models: (1) models where one considers a planet as one of the final cooled pieces of the exploded star or super-nova, (2) models where one considers a planet as a result of the cosmic dust condensation.

Relative to any model from the first group, it is impossible to distinguish in any piece of the terrestrial mass genetically real parent and daughter nuclei from the admixture of the same kinds of nuclei formed in other parts of the cooled and transformed parent stellar piece. Therefore one can approximately suppose that the earth age is the sum of the parent star or super-nova age before exploding and of the consequent age of the formed earth. The last age can be determined also by the methods of the nuclear chronometry but in different ways inside the earth and on the surface of the earth. Deeply inside the earth one has to consider the consequences of the formation and decay of the excited nuclei-chronometers during the preceding stellar (and super-nova) nucleosynthesis and during subsequent planet cooling in the melted magma (inside the earth). And in the surface layers of the earth (up to the depth of  $\sim 5$ meters) we can consider the influence of the cosmic radiation which was presented above. For both cases it is also necessary to take into account the unknown now initial nonzero quantity of the daughter nuclei in the earth (in the examined earth pieces) from the previous stellar (nucleosynthesis and chronometer-decay-chain) processes.

Relative to any model from the second group, from the very beginning it is necessary to take into account the cosmic dust origin. Hypothetically the cosmic dust was born partially simultaneously with first stars after Big Bang and partially during the star evolution - from the cooled micro-rejections out of stars and super-nova during their perturbations and explosions. And now there is neither a systematic theory of the dust origin, independent from the star origin, nor a systematic theory of the dust condensation $\rightarrow$ clotting into a planet. We can approximately evaluate the mean existence time of the earth beginning from the hypothetical mean instant of the conventional dense clotting of the condensed dust into the planet by the methods of nuclear chronometry if we know the real initial quantities of the parent and the daughter nuclei just in this mean instant. \emph{And a nonzero initial quantity of the daughter nuclei always leads to a real diminishment of the evaluation of the decay time - a larger diminishment for a larger initial quantity of the daughter nuclei.} Moreover, we have to take into account the constant excitation of radioactive nuclei-chronometers by the cosmic radiation and then, in the case of large masses, also the formation of $\gamma$-emission-absorption chains with accompanied multiple excitations of nuclei-chronometers.


\section{Conclusion
\label{sec.conclusion}}

The presented simplified estimations bring to the conclusion that the usual (non-corrected) ``nuclear  clocks'' do really indicate not to the realistic values but to the \emph{upper limits} of the durations of the $\alpha$-decay stellar and planet  processes and also that the realistic durations of these processes have to be noticeably lesser.

\renewcommand{\refname}{Reference:}

\end{document}